# Transient Measurement of Near-field Thermal Radiation between Macroscopic Objects

**Sen Zhang[1], Yongdi Dang[1], Xinran Li[1], Yuxuan Li[1], Yi Jin[1], Pankaj K Choudhury[1], Jianbing Xu[2] and Yungui Ma[1,*]**

[1]State Key Lab of Modern Optical Instrumentation, Centre for Optical and Electromagnetic Research, College of Optical Science and Engineering; International Research Center for Advanced Photonics, Zhejiang University, China

[2]Department of Electronic Engineering and Materials Science and Technology Research Center, Chinese University of Hong Kong, Shatin, Hong Kong, China

**Abstract:** The involvement of evanescent waves in the near-field regime could greatly enhance the spontaneous thermal radiation, offering a unique opportunity to study nanoscale photon-phonon interaction. However, accurately characterizing this subtle phenomenon is very challenging. This paper proposes a transient all-optical method for rapidly characterizing near-field radiative heat transfer (NFRHT) between macroscopic objects, using the first law of thermodynamics. Significantly, a full measurement at a fixed gap distance is completed within tens of seconds. By simplifying the configuration, the transient all-optical method achieves high measurement accuracy and reliable reproducibility. The proposed method can effectively analyze the NFRHT in various material systems, including $SiO_2$, SiC, and Si, which involve different phonon or plasmon polaritons. Experimental observations demonstrate significant super-Planckian radiation, which arises from the near-field coupling of bounded surface modes. Furthermore, the method achieves excellent agreement with theory, with a minimal discrepancy of less than 2.7% across a wide temperature range. This wireless method could accurately characterize the NFRHT for objects with different sizes or optical properties, enabling the exploration of both fundamental interests and practical applications.

**Key words:** Near-field radiative heat transfer, surface mode, all-optical, nanophotonics

## 1. Introduction

Thermal radiation is a ubiquitous phenomenon associated with the emission of oscillating polar molecules or charged particles in matter. When the gap distance between the emitter and the receiver is far smaller than the thermal wavelength ($\lambda_T \approx 10$ μm at room temperature), thermally excited evanescent waves could tunnel through the gaps and significantly strengthen the heat flux. It becomes more obvious when assisted by bounded surface modes such as surface plasmon polaritons (SPPs),[1-3] surface phonon polaritons (SPhPs)[4-6] or hyperbolic modes.[7-10] Henceforth, engineering the emission surface to have a high local density of states (LDOS) of photons is a compelling area of research for controlling thermal photons and creating new applications.[11, 12] There are abundant interesting physical phenomena associated with the near-field optics, which have great potential for various applications such as thermophotovoltaic cells,[13-17] thermal logic circuits,[18-21] solid-state refrigeration,[22, 23] thermal management systems.[24-26]

For these purposes, sophisticated measurements for near-field thermal flux are very important. Over the past two decades, experimental skills on NFRHT have been substantially developed for different material systems[27-34] in different configurations including tip-plate,[13, 23, 35, 36] tip-membrane[30] and plate-plate.[4, 5, 27-29, 37, 38] Most of these methods need electric contacts for the functions of heating and sensing. In many cases, the part of heat conducted along electric wires cannot be ignored, in particular for the setup measuring the bulks,[39] which will complicate the calibration process. To partially solve the issue, wireless optical methods have been developed in near-field thermal transfer systems. For example, Takuya Inoue *et al*. developed a near-field thermophotovoltaic cell that utilized laser and reflectivity for heating and temperature sensing.[40] In our recent work, we demonstrated an all-optical approach to characterize the NFRHT between the graphene-SiC heterostructures utilizing laser to heat the emitter and thermography to measure temperature.[41] With the fixed gap distance design, the sample holder for the emitter was also saved, which further helped to simplify the measurement. However, these were still a thermostatic measurement. Long measurement duration needs great efforts in controlling the system errors, for example in order to precisely take into account the possible temperature change of the environment.

In this work, we propose a transient all-optical method to rapidly characterize the NFRHT between macroscopic surfaces. Different from the previous thermostatic measurement, the temperature cooling process after turning off the heating laser was recorded by the thermography and utilized to analyze the near-field heat flux. This method is based on the condition that the emitter has a negligible temperature gradient along the thickness as the equivalent radiation conductance is in orders far smaller than the solid thermal conductance of the emitter. In terms of the principle of energy conservation, the transient heat flux was derived from the temperature reduction of the emitter regarded as a thermal reservoir. The transient measurement greatly shortens the experiment time period. The reliability of the method was inspected by measuring the NFRHT between semi-finite surfaces made of different matters (SiC, SiO$_2$, and Si). The results indicate a close correspondence between the experimental measurements and theoretical predictions, with a minor discrepancy of less than 2.7% across a wide temperature range. The applicability of the method is also briefly discussed.

## 2. Method

### 2.1 Material and theoretical analysis

We apply the methodology to three uniform pairs made of semi-infinite materials, namely SiC-SiO$_2$, SiO$_2$-SiO$_2$ and Si-SiO$_2$. These samples show different coupling modes in the near-field due to their distinct phonon or plasmon resonances in infrared bands. Drude model $\varepsilon_{Si} = \varepsilon_\infty - \omega_{p,e}^2/(\omega^2 + i\gamma_e\omega)$ is used to describe the dielectric response of the n-type Si with the infinite frequency $\varepsilon_\infty = 11.67$, plasma frequency $\omega_{p,e} = 5.6\times10^{14}$ rad/s and collision frequency $\gamma_e = 1.1\times10^{14}$ rad/s.[28, 37] The dielectric property of SiC is fitted by a Lorentz model $\varepsilon_{SiC} = \varepsilon_\infty (\omega^2 - \omega_{LO}^2 + i\omega\gamma)/(\omega^2 - \omega_{TO}^2 + i\omega\gamma)$ with $\varepsilon_\infty = 6.7$, longitudinal optical phonon frequency $\omega_{LO} = 1.8\times10^{14}$ rad/s, transverse optical phonon frequency $\omega_{TO} = 1.5\times10^{14}$ rad/s and damping factor $\gamma = 9.0\times10^{11}$ rad/s.[41] The dielectric function of SiO$_2$ is numerically fitted using $\varepsilon(\eta) = \varepsilon_\infty + \sum_j[g_{cj}^{kkg}(\eta) + ig_{cj}(\eta)]$ with the coefficients $\varepsilon_\infty$, $g_{cj}^{kkg}(\eta)$ and $g_{cj}(\eta)$ could be found elsewhere.[42] Fig. 1a shows the permittivity spectra of the three materials examined from their reflection spectra by a Fourier transform infrared spectroscopy (FTIR, Vertex 70, Bruker, Germany). SiC and SiO$_2$ samples exhibit prominent optical phonon resonances in the long infrared range, whereas weakly doped Si only displays a simple lossy dielectric.

Theoretical analysis on NFRHT is conducted first to understand the underlying physics. Fig. 1b shows the near-field heat flux versus gap distance. The results indicate that the most significant enhancement in heat flux can be attributed to the identical SPhPs resonances between $SiO_2$-$SiO_2$. On the other hand, the heat flux between SiC-$SiO_2$ is smaller than that of Si-$SiO_2$, mainly due to a mode frequency mismatch between $SiO_2$ SPhPs and Si plasmon. The mode frequency distributions between different materials are explored in Fig. 1c, which shows the calculated near-field spectral heat flux, $[\Theta(\omega, T_1) - \Theta(\omega, T_2)]H(\omega, d)$, where $\Theta$ represents the mean energy of the Planckian oscillator and $H(\omega,d)$ is the heat transfer coefficient dependent on angular frequency $\omega$ and gap distance $d$.[37] The analysis is conducted on SiC-$SiO_2$, $SiO_2$-$SiO_2$ and Si-$SiO_2$ configurations at $d$ = 100 nm, $T_1$ = 373 K (emitter's temperature) and $T_2$ = 303 K (receiver's temperature). In Fig. 1c, for the $SiO_2$-$SiO_2$ pair, the spectral heat flux exhibits two prominent oscillation peaks around $0.9\times10^{14}$ and $2.2\times10^{14}$ rad/s associated with their SPhPs frequencies. The SiC-$SiO_2$ pair displays only one peak around $1.78\times10^{14}$ rad/s within its reststrahlen band. Additionally, the Si-$SiO_2$ combination also supports two minor spectral flux peaks generated by the phonon resonance of $SiO_2$. The heat flux magnitude is determined by overlap degree of dispersion bands of the bounded surface modes excited by the emitter and receiver. It is worth mentioning that the near-field radiation for the different combinations is substantially enhanced compared to blackbody radiation (black line) due to the excitation and coupling of surface modes.[43]

In the near-field region, due to the prominent role of surface polaritons, the $p$-polarized waves will dominate in the heat transfer process and in this case, the contribution of $s$-polarization could be neglected. The transmission coefficients in Fig. 1d–f are drawn as a function of the in-plane wave vector $k_\parallel$ and angular frequency $\omega$. Fig. 1d shows that the original SPhPs resonance of $SiO_2$ are weakened, while another mode associated with the SiC phonon polariton is excited around $1.78\times10^{14}$ rad/s. Inside the light cone (travelling waves), in contrast, the transmission coefficients at the SiC SPhPs frequencies are largely suppressed due to the phonon resonance of SiC. In Fig. 1e, two prominent transmission bands appear around $0.9\times10^{14}$ and $2.2\times10^{14}$ rad/s, indicating a strong coupling between identical $SiO_2$. In Fig. 1f, for Si-$SiO_2$, the slightly-doped Si cannot provide adequate optical loss and SPPs at thermal infrared band, therefore in this situation the tunneling bands arising from the $SiO_2$ SPhPs are weakened. These figures imply the mechanism of near-field coupling strength mediated by the mode distributions of different material combinations. Thorough validation of our method can be achieved by measurements of NFRHT via various coupling modes.

## 2.2 Measurement setup

Fig. 2a gives the schematic of our all-optical transient NFRHT measurement conducted inside a high vacuum chamber (< $10^{-6}$ Torr). An infrared window is used to transmit 532 nm heating laser and spontaneous emission from the emitter. A ZnSe lens (diameter 10 mm and focal length 25.4 mm) is used to focus the input laser beam which reversely behaves as the front side element of a $4f$ radiative thermal image optical system. With these lenses, the spatial resolution of our thermograph could reach ~ 50 μm, which is adequate for reading the local temperature of our millimeter-sized emitter (surface area 1 × 1 $mm^2$). A thin blackbody paint with emissivity ~1 is coated on the top surface of the emitter to absorb the laser light and emit thermal photons for temperature measurement. For the near-field samples, the emitters and receivers are three pairs of homogenous natural materials, i.e., SiC-$SiO_2$, $SiO_2$-$SiO_2$ and Si-$SiO_2$, spaced by four cylindrical photoresist nanopillars (SU8 2000.5, thermal conductivity = 0.3 $Wm^{-1}K^{-1}$) with a radius of 1.3 μm. On the receiving surface, array of photoresist structures made of four cylinders and a $1.3 \times 1.3$mm$^2$ square mark (green portions in Fig. 2a) are fabricated by photolithography (MA6/BA6, SUSS MicroTec, Germany). The two components are appropriately aligned with the help of photoresist marks. The gap distances are reliant on the heights of spacing nanopillars, which could be regulated within 100-600 nm by modifying the fabrication recipes, specifically the spin coating speed and dilution ratio, as adopted in previous work.[41] To guarantee parallelism between the emitter and the receiver, the bending conditions is analyzed using a profilometer (Veeco NT9100, America), as shown in Fig. 2b. As a result, the bending of each sample is approximately 20 nm across nearly $1 \times 1$ mm$^2$ area, which is far smaller than the gap distances applied in the subsequent measurements. Note that a hard bake pretreatment is executed to improve mechanical stability.

In this study, analysis on cooling process of emitter is employed to determine the near-field heat flux. The experimental procedure first involves heating the emitter by a laser, which is then switched off until the temperature reaches the set-point. In the near-field situation, the emitters usually cool down from ~100 °C to near ambient temperature within tens of seconds, much shorter compared to general thermostatic measurements.[37, 39] To record the real-time temperature of the cooling emitter, a thermograph (HPM 11, HIKIMICRO, China) is used with a frame rate of 25 Hz, which is sufficient for most NFRHT measurements. The relation between the exact temperature $T_e$ and the thermograph temperature $T_t$ is calibrated *in situ* using a thermal resistance and a thermograph. A linear correlation ($T_e = 1.11 \times T_t$-7.83 (°C)) is detected within the measurement range of thermograph (0~150 °C). The calibration results yield a temperature uncertainty comparable to that of the electrical method, with an approximate range of ±0.1 K. The changes in stored thermal energy of the emitter are subsequently documented and converted into energy dissipations occurring through three distinct channels. In this study, it is assumed that the emitter with a thickness of 0.5 mm has a homogeneous temperature throughout its volume. Note this assumption may not be valid for extremely small gap distances, such as 10 nm or smaller,[44] corresponding to very strong NFRHT, which is not the case conducted in the current experiment working on macroscale emission surfaces. The thermal capacitive time constant for the emitter is about 30s, which is far larger than the diffusion time (~ 0.3 μs) across the thickness.[45] Consequently, there only exists a tiny temperature gradient between the top and bottom surfaces of the emitter under large heat flux (< 0.75 K for 2000 W/m$^2$), which has negligible effects on the NFRHT. Similarly, the temperature difference between the top and bottom surfaces of the receiver in our setup is much less and could be practically neglected in our analysis.

## 3. Results

### 3.1 Thermal circuit and calibrations

The thermal circuit of our system is illustrated in the inset of Fig. 2a. The energy stored in the emitter (temperature $T_1$) is dissipated through three pathways: (i) far-field radiation into the chamber ambience (temperature $T_{am}$), (ii) conduction along the nanopillars, and (iii) near-field radiation to the receiver (temperature $T_2$). Based on the uniform temperature assumption, the transient output thermal power $Q_{out}(t)$ of the emitter can be calculated by differentiating the measured transient temperature over time, applying the principles of energy conservation

$$Q_{out}(t) \equiv c_i m_i dT_1/dt \qquad (1)$$

where the variables $c$ and $m$ are the specific heat capacity and mass, respectively, and the subscript $i$ (= SiO$_2$, SiC, Si) represents the different material selection for the emitter used here. The specific heat capacities and masses for SiO$_2$, SiC and Si are 0.74 J/(g·K) and 1.1 mg, 0.71 J/(g·K) and 1.5 mg and 0.70 J/(g·K) and 1.2 mg, respectively. In the experiment, there is a tiny temperature gradient along the thickness direction of the emitter. The top surface temperature is adopted in calculating the release of heat energy stored inside the emitter. The approximation is valid only if the thermal resistance of the vacuum channel is in orders far larger than the emitter as satisfied by our experiment. Based on the first law of thermodynamics, we have

$$Q_{\text{out}}(t) = Q_{\text{FF}}(t) + Q_{\text{con}}(t) + Q_{\text{NF}}(t) \qquad (2)$$

where $Q_{\text{FF}}$, $Q_{\text{con}}$ and $Q_{\text{NF}}$ represent the transient thermal power dissipated through the far-field radiation, conduction and near-field radiation channels, respectively. According to their characteristics, we use the following formulas to estimate these quantities[28, 41]

$$Q_{\text{FF}} = A_{\text{FF}} \times S_{\text{FF}}\sigma(T_1^4 - T_2^4) \qquad (3)$$

$$Q_{\text{con}} = A_{\text{con}} \times \frac{S_{\text{con}}k(T_1 - T_2)}{d} \qquad (4)$$

$$Q_{\text{NF}} = A_{\text{NF}} \times \frac{1}{4\pi^2}\int_0^\infty [\Theta(\omega, T_1) - \Theta(\omega, T_2)]\, H(\omega, d)\mathrm{d}\omega \qquad (5)$$

In the above, $A_{\text{FF}}$, $A_{\text{NF}}$ and $A_{\text{con}}$ are the area of the three heat dissipation channels, $\sigma$ and $k$ are the Stefan-Boltzmann constant and thermal conductivity (0.3 Wm$^{-1}$K$^{-1}$), $\Theta$ is mean energy of Planckian oscillator and $H(\omega,d)$ is the heat transfer coefficient. The coefficient $S_{\text{FF}}$ estimates the effective emissivity of the emitter and chamber environment and $S_{\text{con}}$ is a correction factor for the conductance of spacing nanopillars due to the existence of contact thermal resistance.[46] In this work, the temperature $T_2$ of the receiver is equal to that of the ambient. Substituting eqn (2)–(5) into eqn (1) and subtracting the far-field radiation and conduction power, one can calculate the component of the near-field radiation power from the measurement.

The calibration was first conducted to evaluate the far-field radiation and effective conductance, namely to obtain the coefficients $S_{\text{FF}}$ and $S_{\text{con}}$. Three different pairs of substrates: SiC-Al/SiO$_2$, SiO$_2$-Al/SiO$_2$ and Si-Al/SiO$_2$ are used in experiments. A 150 nm thick aluminum film was sputtered on the surface of the SiO$_2$ receiver to minimize the NFRHT, which was not considered in the calibration due to its comparatively small magnitude. The photoresist nanopillars were fabricated on the receiver side to ensure the same contact conditions between emitters and pillars. At least three identical measurements were replicated to check the reproducibility. Calibration procedures were implemented at near 150 nm gap distances to increase overall dissipation power and improve calibration accuracy. The recorded cooling curves (circles) for the three pairs are sparsely plotted in Fig. 3a-c for clarity. The transient output energy $Q_{\text{out}}(t)$ of the emitter shown in the lower panels, was derived from the cooling curves using eqn (1) at $Q_{\text{NF}} = 0$. The quantities $S_{\text{FF}}$ and $S_{\text{con}}$, which have different temperature dependences (quadric or linear relation for the far-field radiation or thermal conduction channels), can be obtained by fitting the transient output function $Q_{\text{out}}(t)$ using eqn (1) over a relatively wide temperature range. For the SiC-Al/SiO$_2$, SiO$_2$-Al/SiO$_2$ and Si-Al/SiO$_2$ pairs, the fitted coefficients ($S_{\text{FF}}$, $S_{\text{con}}$) are (0.20±0.02, 0.69±0.03), (0.22±0.01, 0.77±0.05) and (0.26±0.02, 0.47±0.03), respectively. These coefficients are determined based on repeated measurements, and the uncertainties are treated as coefficients errors. The fitted results show high accuracy compared to the experimental data. The far-field radiation coefficients $S_{FF}$ for the three pairs range from 0.20 to 0.26, indicating limited heat transfer between the emitter and the chamber, with the deviations attributed to the differences in emissivity of each material. On the other hand, the conduction coefficient $S_{con}$ varies within a relatively large range of 0.47 to 0.77. This is expected as the thermal contact resistance depends on factors such as topology, materials, and pressure.[46] This suggests that the real thermal contact resistance is sensitive to the species contacting the nanopillars. The measured coefficients ($S_{FF}$, $S_{con}$) are directly used in the subsequent NFRHT measurement.

## 3.2 Measurement on SiC-SiO₂ at same gap distance

The transient all-optical method is examined by SiC-SiO₂ sample pairs at the same gap distance firstly. Three SiC emitters were placed on a SiO₂ substrate to ensure uniformity, with precise control over the contact and ambient conditions. Fig. 4 demonstrates the cooling process and the retrieved near-field radiation data under a gap distance of 235 nm. Note here the $S_{FF}$ and $S_{con}$ quantities are obtained from initial calibration and the error bars are calculated considering the coefficient errors. The significant consistency is demonstrated by three replication measurements and quantified by the normalized (normalized by the mean value of each data point) mean absolute error (MAE) which is less than 7.3% overall. Furthermore, it is evident that the experimental results match well with the theoretical predictions. The relatively small amplitude is ascribed to the mode mismatching between SiC and SiO₂. These results also indicate the robustness of our measurement method.

## 3.3 Measurement on three pairs at different gap distances

Next, we apply the rapid transient all-optical method to three different pairs to demonstrate the applicability of detecting NFRHT with diverse mode couplings. The measured cooling curves for the three samples of SiC-SiO₂, SiO₂-SiO₂ and Si-SiO₂ are plotted in Fig. 5a–c, respectively. The process only takes about 1 minute for $T_1$ to reduce from 100 °C to near ambient, which is significantly faster compared to traditional steady measurements.[28] The gap distances for the pairs of the SiC-SiO₂, SiO₂-SiO₂ and Si-SiO₂ samples are 202 nm, 515 nm and 400 nm, respectively. These measurements are performed under three different gap distances to assess the applicability and accuracy of this method. By utilizing the correction coefficients ($S_{FF}$, $S_{con}$) measured above (see Fig. 3), we could derive the transient heat flux for three different channels. Fig. 5d–f depict the retrieved near-field heat flux component with the temperature difference $\Delta T$ between the emitter and receiver. Here $Q_{NF\ exp}$ represents the measured data and $Q_{NF\ cal}$ represents the theoretical results obtained using the thermal fluctuation dissipation theory, considering gap uncertainties.[47] The comparison between theory and measurement shows good agreement. The measured near-field heat flux is 2-3 times stronger than the blackbody, highlighting the prominent role of evanescent waves in NFRHT. We evaluate the accuracy of the methodology using the absolute deviation ratio, defined as $|Q_{NF\ exp}/Q_{NF\ cal}\ -1|\times100\%$ (triangle symbols in Fig. 5) between the measured and calculated data. For the SiC-SiO₂ sample (Fig. 5d), $\Delta Q$ fluctuates with a small magnitude of less than 10% at $\Delta T > 25$°C. For the SiO₂-SiO₂ sample (Fig. 5e), $\Delta Q$ is smaller than 2.7% over entire measured temperature range ($18°C<\Delta T<46°C$). For the Si-SiO$_2$ sample (Fig. 5f), $\Delta Q$ is less than 4.6% for $\Delta T >25°C$. To ensure high accuracy, it is crucial to control and characterize the transient physical state of the sample. During measurement, geometry deformation caused by large temperature reduction may be the major issue affecting the accuracy, which can explain the relatively large $\Delta Q$ fluctuation at small $\Delta T$. The difference in accuracy among three samples may be attributed to the different proportions of near-field radiation among total dissipation channels. The near-field radiation proportion for SiC-SiO$_2$ (around 40%) is nearly 2/3 compared to the other two measurements (around 60%), leading to a poor accuracy. In addition, the error bars in Fig. 5 could be normalized and the average quantities for SiC, SiO$_2$ and Si are $\pm9.8\%$, $\pm3.5\%$ and $\pm4.8\%$, respectively. For the less coupled SiC-SiO$_2$ sample, the proportion of near-field radiation is small, resulting in a relatively larger influence of coefficient errors ($\pm9.8\%$). The accuracy achieved here is relatively high as typical electrical-based methods unavoidably introduce a system uncertainty of around $\pm5\%$ ascribed to the inherent uncertainties in electrical components such as heat flux meter.[28, 37, 48]

## 4. Discussion

The wireless transient measurement adopted above could effectively simplify the calibration and measurement process compared with a traditional long-time thermostatic method. This is achieved through a simple experimental setup and the use of optical heating and sensing techniques. The system errors in typical electrical-based methods mainly come from the uneven temperature distribution due to adhesive and multilayer structures as well as the accuracy of heat flux meter or other electrical components.[48] This requires time-consuming and difficult thermostatic calibrations on temperature and heat flux. However, our method focuses on the correction of far-field and conductive energy transfer of the emitter, which only requires a metal layer deposited on the receiver side. Optical heating and sensing techniques are widely used in optics, but precise temperature measurement for a millimeter surface using a thermograph is usually challenging. The utilization of a ZnSe lens in our method helps to magnify the image and enhance the resolution.[49] Additionally, maintaining the parallelism between emission surfaces is easier with our method as the emitter sustains minimal extra stress during the cooling or heating process. Traditional electric measurements, relying on adhered heaters and sensors, may unwantedly distort the surface flatness with large temperature changes due to thermal expansion mismatch. The transient technique is particularly useful for describing the NFRHT for small objects at fixed gap distances, such as samples made of metasurfaces[50, 51] or two-dimensional materials[52] that are difficult to manufacture in larger sizes. For these cases, to derive the released heat power, it shall be enough to know the specific heat capacitance of the thick substrates holding these samples. In this study, we demonstrate the feasibility of applying this technique. The current experimental setup can be further improved, for example, by replacing thermography with temperature sensitive photoluminescence measurement to achieve higher resolution.[53]

## 5. Conclusion

An all-optical transient measurement method is proposed to characterize the NFRHT between macroscopic objects. The validity and efficiency of this method are examined on three uniform samples with different near-field modes coupling strengths. Initially, the method is applied to three identical SiC emitters at the same gap distance and the reproducibility is examined to be less than 7.3%. Subsequently, the method is demonstrated on three different samples at gap distances ranging from 200 to 500 nm. The measured results show good agreement with the theory, with a minimum deviation ratio of less than 2.7% over a wide temperature range. A noticeable advantage of this method is that it utilizes a wireless measurement configuration with a simplified emitter structure, which helps to reduce the calibration process and measurement duration. The proposed transient NFRHT measurement method could be extended to characterize near-field thermophotonic interactions of smaller objects fabricated using advanced nano-fabrication techniques or two-dimensional materials with exotic infrared features.

## Acknowledgement
YGM thanks the partial supports from National Natural Science Foundation of China 62075196, Natural Science Foundation of Zhejiang Province LXZ22F050001 and DT23F050006, and Fundamental Research Funds for the Central University. JBX would like to thank RGC for support via AoE/P-701/20.

## Data availability
The data that support the findings of this study are available from the corresponding author upon reasonable request.

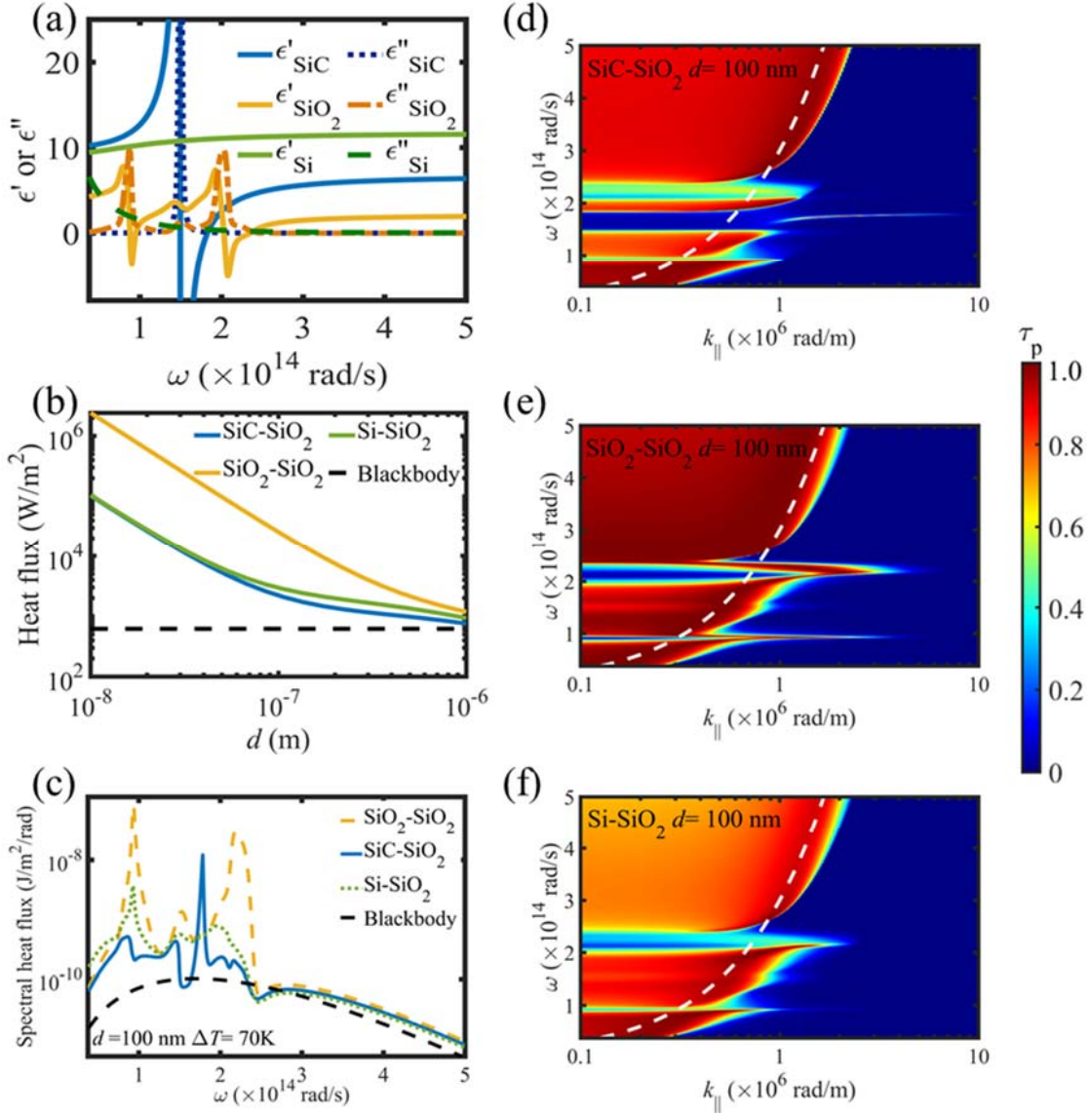

**Fig. 1** Theoretical analyses of the NFRHT. The calculating conditions: $T_1 = 373$ K, $T_2 = 303$ K and $d = 100$ nm. (a) Dielectric spectra of three materials: SiC, SiO$_2$ and Si. $\varepsilon'$ and $\varepsilon''$ represent the real and imaginary parts of the dielectric constant, respectively. (b) NFRHT at different gap distances. (c) Spectral heat flux of SiO$_2$-SiO$_2$, SiC-SiO$_2$, Si-SiO$_2$ and blackbody. (d)-(f) Transmission coefficients ($\tau_p$) for $p$-polarization as a function of in-plane wavevector $k_\parallel$ and angular frequency $\omega$. The white dashed line profiles the light cone edge in vacuum.

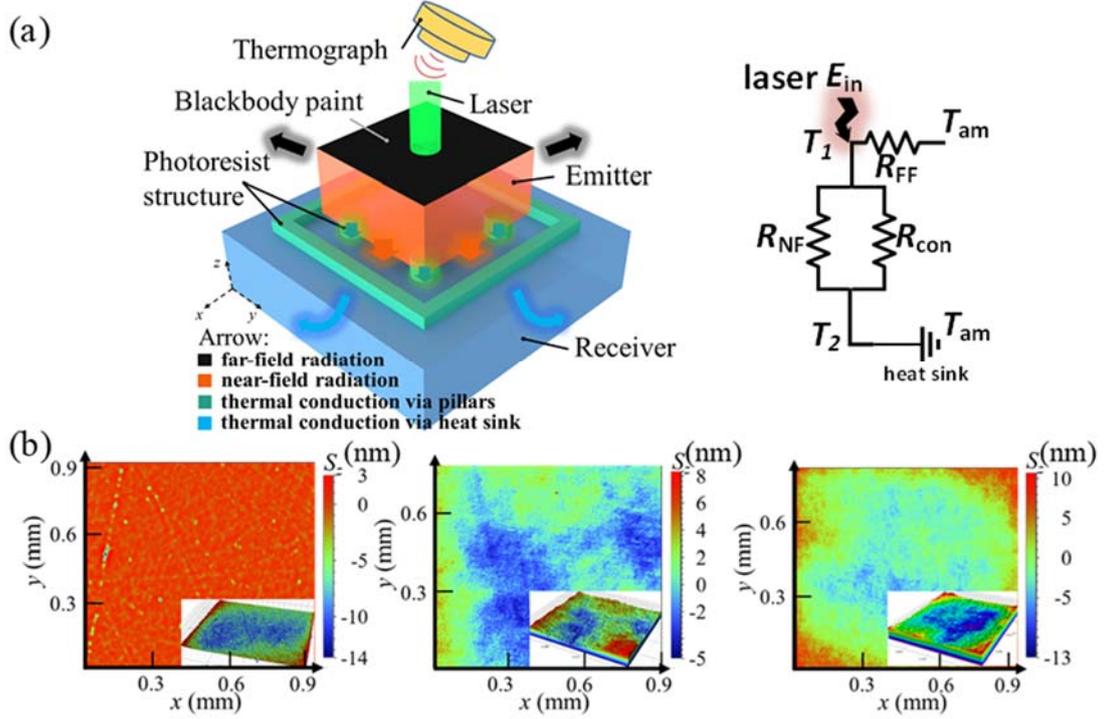

**Fig. 2** (a) Schematic diagram of the NFRHT measurement configuration and simplified thermal circuit. The thermal circuit shows the thermal energy dissipation paths. $E_{in}$ is the input of the thermal energy from laser source, and $R_{FF}$, $R_{NF}$ and $R_{con}$ represent the effective thermal resistances of the far-field radiation, near-field radiation and conduction, respectively. $T_1$, $T_2$ and $T_{am}$ ($=T_2$) are the temperature of emitter, receiver and the ambient environment, respectively. (b) Bending contour diagrams of three samples: SiC, SiO$_2$ and Si, as shown from left to right, with measurement areas of 0.827, 0.701 and 0.799 mm$^2$, respectively. $S_z$ is the bending conditions of surfaces in the unit of nanometer. Note that the insets are the re-build topographies based on main figures.

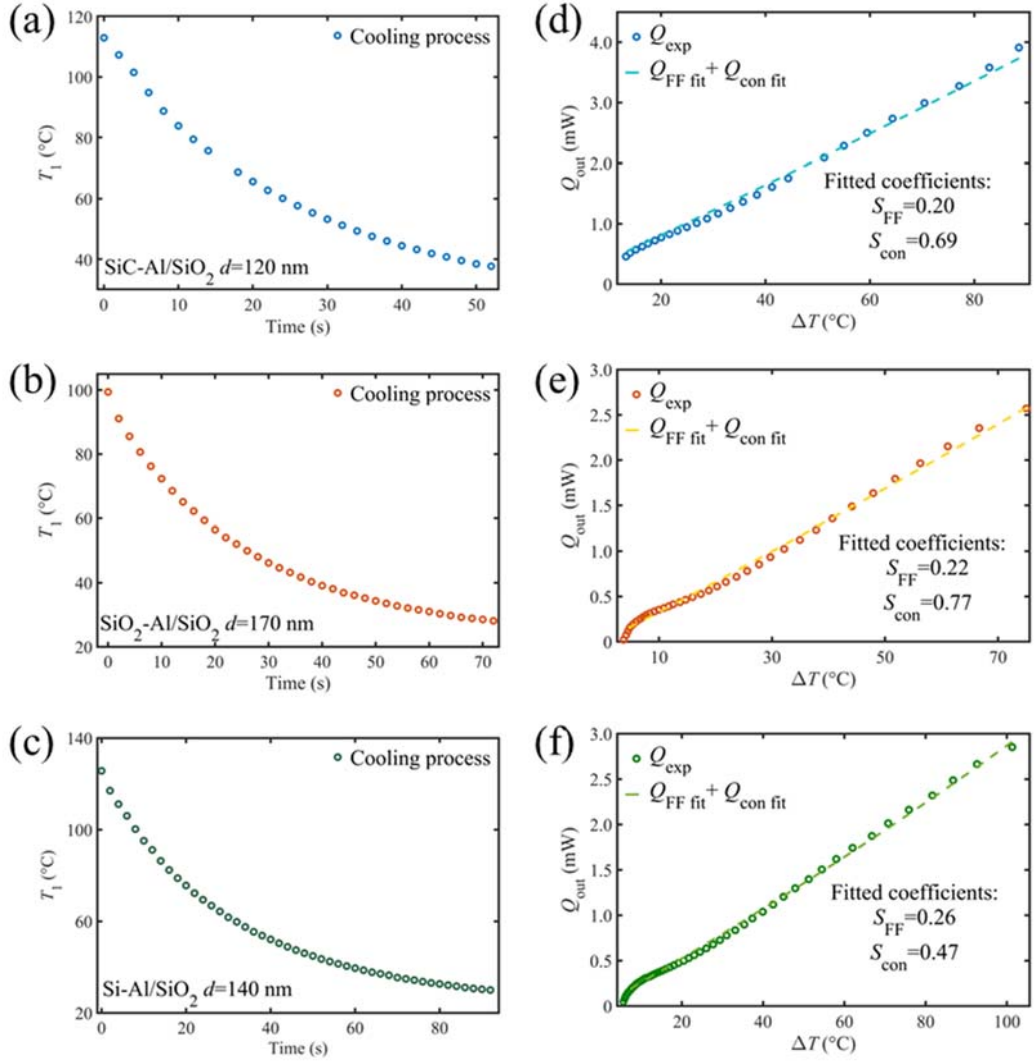

**Fig. 3** Experimental calibrations for rapid transient method on three homogenous samples made of SiC-Al/SiO$_2$, SiO$_2$-Al/SiO$_2$ and Si-Al/SiO$_2$. The cooling processes of these three samples are plotted in (a)-(c) while retrieved experimental dissipation power data ($Q_{exp}$) and fitting curves ($Q_{FF\,fit}$ + $Q_{con\,fit}$) are plotted in (d)-(f) with dots and dashed lines, respectively.

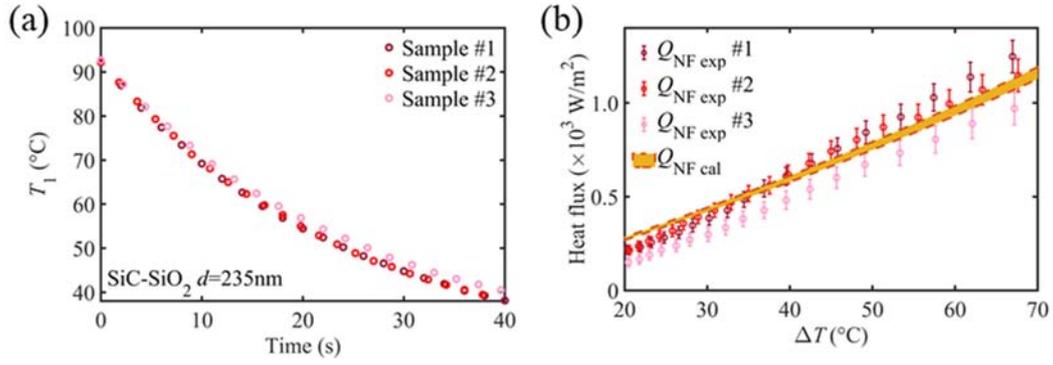

**Fig. 4** Measurements of SiC-SiO$_2$ at 235 nm gap distance. (a) Cooling processes for three identical SiC-SiO$_2$ samples, denoted as #1, #2 and #3. (b) Retrieved experimental near-field radiation data ($Q_{NF\ exp}$) for sample #1, #2 and #3. The error bars are calculated based on the coefficient calibration. Theoretical predictions ($Q_{NF\ cal}$) are plotted as yellow area considering the bending of chips as depicted in Fig. 2b.

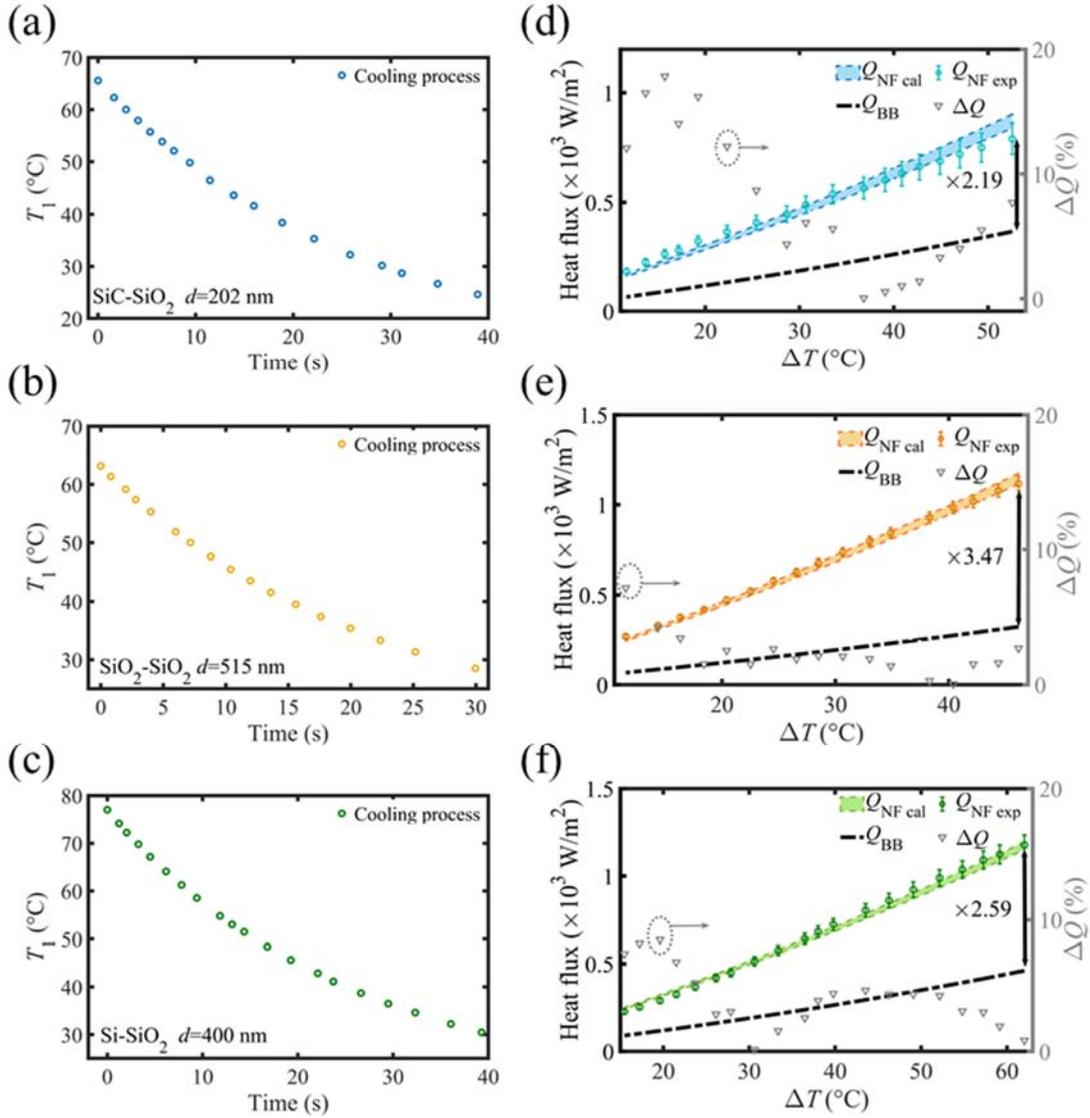

**Fig. 5** Measured results of the NFRHT for three homogenous samples at three gap distances. (a)–(c) Cooling processes of emitters for three homogenous samples of SiC-SiO$_2$, SiO$_2$-SiO$_2$ and Si-SiO$_2$, respectively. (d)–(f) Retrieved experimental near-field radiation data $Q_{NF\ exp}$ (donuts symbols with error bars calculated based on coefficients calibrations) and theoretical predictions ($Q_{NF\ cal}$, shadow area considering the bending of chips as depicted in Fig. 1b) of the three samples. The curves for the blackbody limit ($Q_{BB}$, dashed lines) are also given here for comparison, from which, we estimate the enhancement factors of near-field radiation. The absolute deviation ratios $\Delta Q$ (inverted triangles) are also plotted for the right *y*-axis in (d)–(f).